A Seminar Report submitted in partial fulfilment
of the requirements for the award of degree

Integrated Dual Degree,

B.Tech In CSE & M.Tech In IT

# Techniques for Deep Query Understanding

Submitted by

Abhay Prakash

(10211002)

Under the guidance of

Dr. Dhaval Patel

Assistant Professor

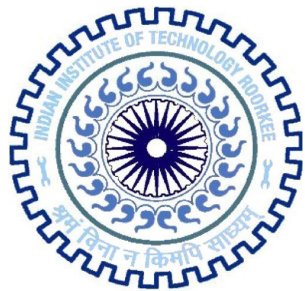

DEPARTMENT OF COMPUTER SCIENCE AND ENGINEERING

INDIAN INSTITUTE OF TECHNOLOGY

ROORKEE – 247667

# ABSTRACT

*Query Understanding* concerns about inferring the precise intent of search by the user with his formulated query, which is challenging because the queries are often very short and ambiguous. The report discusses the various kind of queries that can be put to a Search Engine and illustrates the *Role of Query Understanding* for return of relevant results. With different advances in techniques for deep understanding of queries as well as documents, the Search Technology has witnessed three major era. A lot of interesting real world examples have been used to illustrate the role of Query Understanding in each of them.

The Query Understanding Module is responsible to correct the mistakes done by user in the query, to guide him in formulation of query with precise intent, and to precisely infer the intent of the user query. The report describes the complete architecture to handle aforementioned three tasks, and then discusses basic as well as recent advanced techniques for each of the component, through appropriate papers from reputed conferences and journals.

# TABLE OF CONTENTS

# LIST OF FIGURES

# LIST OF TABLES

# 1 INTRODUCTION

## 1.1 MOTIVATION

In past 20 years, the way Search Queries are interpreted, processed and the way the results are shown to the user has completely changed. The process has advanced from mere retrieval based on text matching to the present stage where search results are tried to be obtained on the basis of true semantic understanding of query, along with context, location, time, user's previous short term and long term browsing activity, his community in social networks and a lot many other factors to precisely figure out the intent of the user.

Across the years, even the way search queries are drafted by the users has changed, primarily because of shift from desktop to mobile environment. As for handheld mobile devices, queries are mostly put by voice instead of typing and while speaking users naturally tend to put grammatically correct queries (Natural Language Queries) instead of Keyword Represented Queries. e.g. a voice query "*who is the best classical singer in India*" is more likely than "*best classical singer India*". This idea is supported by the results of the experiments done by [1], in which they found that out of 22 seventh grade students, 35% students framed natural language queries for search, whereas experienced users usually don't do this. This behavior change can be thought to be because of the way search engines earlier used to return match results based on text match. Though Natural Language Queries (following proper grammar) may be ambiguous, but still they are more helpful than tokens to understand the intent of query. Hence the shift to NL Queries can be leveraged to better understand the intent.

Another kind of queries which has emerged because of mobile devices' environment is contextual questions like "*Where can I eat cheese cake right now*". For such queries system does not need to return a globally retrieved list of documents. To generate relevant personalized results, the Query Understanding Module must pass the intent of the query, user location and time to the underlying Answer Generation Module which is responsible to retrieve the results. These information of location, time are not just passed to answer generation module but also used by the Query Understanding Module itself too, to detect the intent along with other information like context of user previous search, user interests etc. Milad et. al in [2] has discussed use of features of location, user demographic features such as age and gender, region features, user short and long history features for query auto completion, but also gives the idea of using the methodology for intent prediction too.

Another class of queries have emerged due to the advancements occurred in the area of Answer Generation where the user is directly provided with the answer to his query, instead of 10 URLs retrieved. Expert systems like 'Wolfram Alpha ™', 'IBM Watson ™', and the major search engines like Bing and Google are targeting this problem and have already started providing direct answers. An examples is shown in Figure 1.1 where the query is *"who is the president of USA"* and the search engines have directly given the answer *"Barack Obama"*.

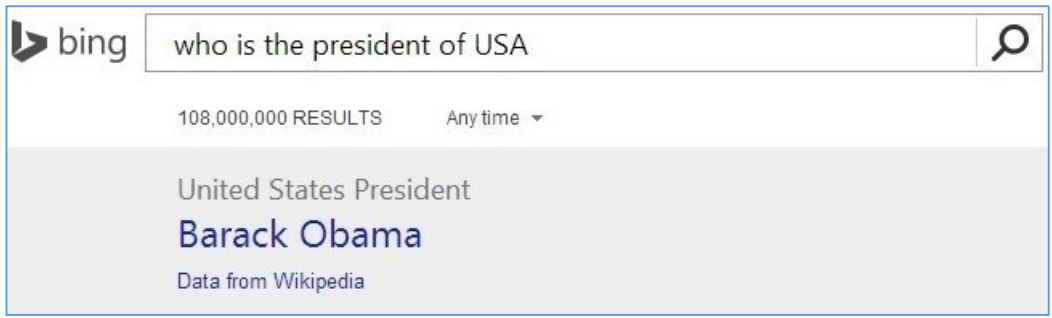

*Figure 1.1: Direct Answering to Query*

With this support to direct answering, a new challenging task has been identified to support Interactive Question Answering, in which the queries are related to the previous ones in the session. An example for this has been shown in Figure 1.2 where each query is dependent on the previous ones in the same session. The four queries are *"who is the president of USA"*, *"what is his age"*, *"who is his wife"* and *"what is her age"*.

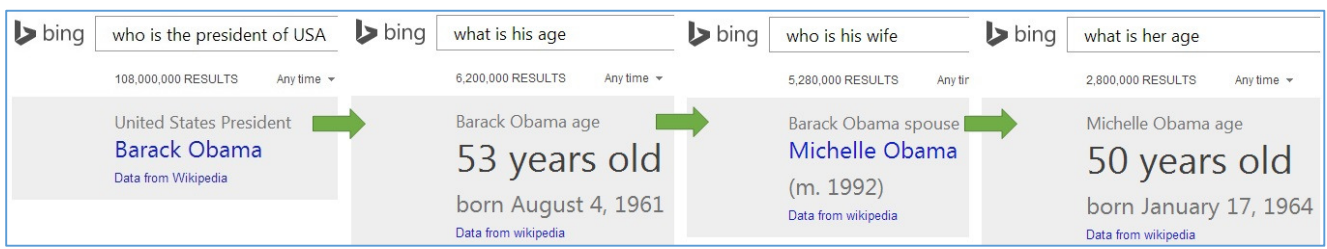

*Figure 1.2: Interactive Question Answering – Sequence of Related Queries put in Bing.*

Note that here the Query Understanding Module must have to consider the context and also take a feedback about the target entity from the Answer Generation Module for previous queries, to understand the query.

Another very interesting and trending use of Query Understanding is in Personal Digital Assistants, which have been recently introduced in various mobile devices. These personal digital assistants are a common interface in which queries/commands are usually given through voice, to execute any kind of operation that can be done on the device like search on internet, search locally on device, schedule meeting, set alarm, making a call or texting a said message to a said contact, playing music from device or internet, or set a reminder for any task etc. Some examples for such Personal Digital Assistants are 'Microsoft Cortana [TM]', 'Apple Siri [TM]' and 'Google Now [TM]'. They have to understand which queries are for web search that they have to redirect to underlying web search engine and which ones are for some local device operation. Some example of queries other than web search queries that can be put in them are like "*make a call to …*", "*set an alarm for …*", "*when I reach home, remind me to …*", "*set a meeting with …*" etc. each of which has to be handled by different application or service running on the device. The Query Understanding Module understands the type of the query and accordingly triggers the corresponding responsible service or fetch result from search engine.

## 1.2 BACKGROUND: QUERY UNDERSTANDING AND 'ADVANCEMENTS IN SEARCH'

Search from the documents and files has been used right from the beginning of Information Technology era – since the knowledge has been started to be digitized. Since then the Search Technology has witnessed three broad phases till now as stated by [3]. These have been described below:

### 1.2.1 Basic Search

In this phase, the search was just based on direct text match and mechanisms like Lucene indexing were used to build inverted index from terms to the documents. The fields (along with text), generally given in form of XML tags like 'Title', 'Heading' etc. were also used to be indexed for restricted retrieval of documents based on user's provided fields in which he wanted to perform search. The documents were used to be submitted manually to a search directory, and the submitter used to provide some additional *'tag'* fields and their values which described the document, with which the searcher may filter out the returned results. This mechanism is still used on nearly all online shopping sites like Flipkart, Amazon etc. where the user filters out the products based on its features, price, manufacturer etc. These fields like price, manufacturer brand, features are known as *facets* and the values that each can take are known as *facet values*. Figure 1.3 shows an example from Flipkart.com.

*Figure 1.3: Facet Example from Flipkart.com*

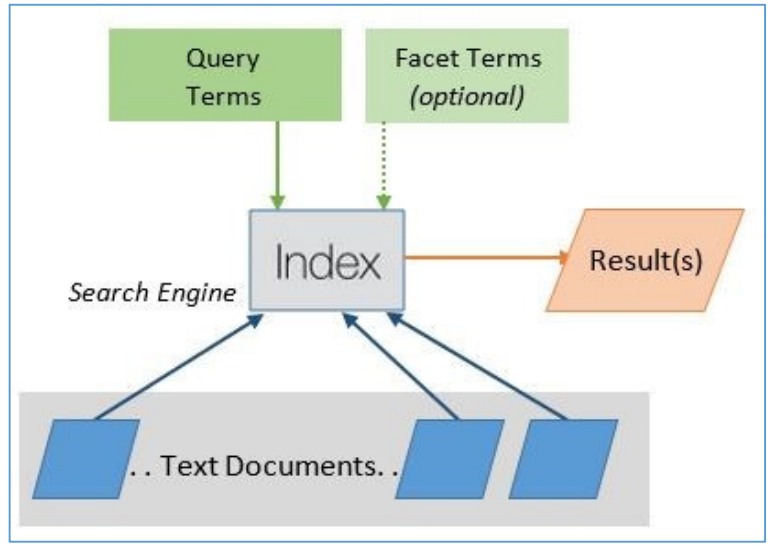

*Figure 1.4: Mechanism in Basic Search*

The queries were similar to Database queries where the facet values were used to restrict the returned results. Figure 1.4 shows the mechanism of a basic search engine, where the query terms are provided and documents containing those words are returned. If facet values are provided for some of the facets then only those documents having the specified facet values for given facets are passed and shown from the returned set of documents. In the search engines of current era, the Query Understanding Module is responsible to understand which facets are being referred to, with what values.

**1.2.2 Advanced Search: Search with Ranking of Results**

This type of search engines used the basic type text search engines at the lowest level but included the concept of *Ranking* the results based on *Relevance* of retrieved documents for the given query. Ranking was used to be done on the basis of TF-IDF and a lot many other signals like keyword relevance, website authority, popularity, link strength etc. It had very basic Query Understanding to identify keywords from the user query. The advancements provided better results than Basic Search as the results were sorted according to relevance (usefulness). Example of such type would be Google as was in its early ages.

**1.2.3 Deep Search: Search with a Deep Understanding**

This type of search technology is the current emerging one. It starves to provide the user with exactly what he wants at that time. Such a system contains a Query Understanding Module that leverages NLP techniques, knowledge of context - user's previous searches and browsing activity, knowledge about his interests, location, time of query, weather at the location etc. to precisely predict the intent and fetch the best results. Such a system must have the following characteristics:

i. *<u>Semantic Understanding for the queries</u>*: This is important for semantic understanding of the queries. Figure 1.5 from [4] shows the level of semantics while matching query to documents. The adjoining examples on each level clarify the scope of semantic understanding of query. e.g. the word *"utube"* has been understood to be *"youtube"* at the word sense level.

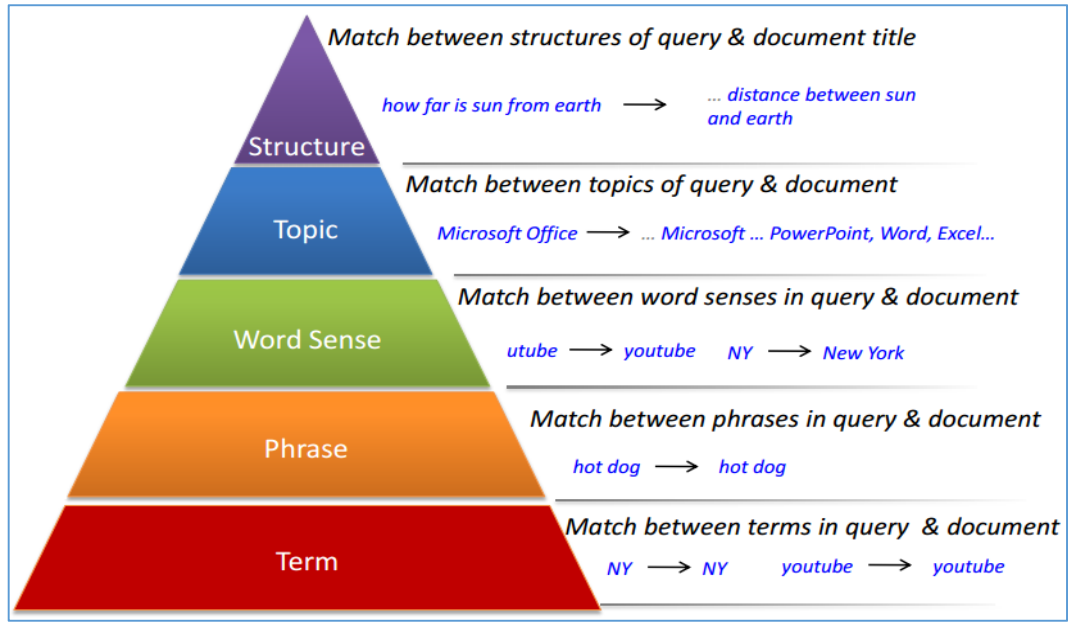

*Figure 1.5: Matching at Different Semantic Levels*

Jingjing Liu et. al in [5] has discussed NLP techniques such as leveraging linguistic parsing knowledge for semantic understanding. This has been discussed in *Section 2.5.1*.

ii. *<u>Context and Previous Tasks Understanding</u>*: Identification of context and building of concept trail can be leveraged to precisely identify the intent in case of ambiguous queries. An example for this could be that suppose a user puts the query *"michael jordan"*, now the intent could be for searching the Basketball player or the well know professor of Machine Learning at Berkley. Now suppose the previous query was *"machine learning"*, it is highly likely that the user's intent is to find the professor Michael Jordan of Berkley. Whereas if the previous queries infer that the context is of sports, then it is more likely that user's intent is to find the famous basketball player. Use of context and walk over the sequence of concepts (concept suffix tree) has been discussed by [6] and [7]. Though queries of a given concept/topic are highly likely to be put consecutively but many sessions contains *interleaved queries* also, as stated by [8]. They found that 30% of sessions contained multiple tasks and 5% of sessions contained interleaved tasks in a sample of 0.5 Billion sessions from Bing. Figure 1.6, taken from [9], shows an example for such a situation. Note that *Reference Query* is the current query put by the user. The session has 4 queries prior to the reference query. Query 1 and 3 are from one task, the same as the task of the reference query. Such previous queries are known as *On-Task Queries*. Whereas the interleaved Queries 2 and 4 are from some different task and hence these are known as *Off-task Queries.*

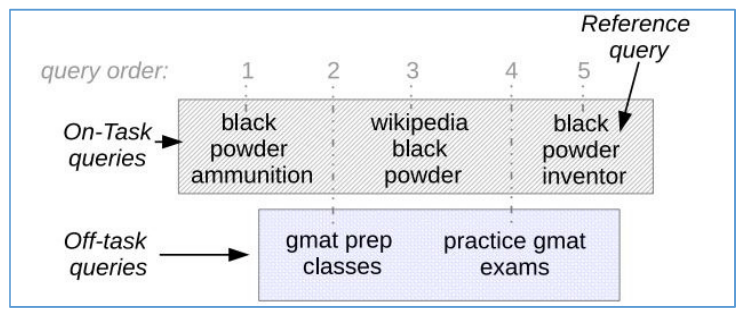

*Figure 1.6: A search context with interleaved tasks*

Allan et. al. in [9] has given a novel approach for identifying the different interleaved tasks and recommending queries based on the identified interleaved tasks. The details on context for Query Understanding have been discussed in *Section 2.2.1* and *Section 2.2.2*.

iii. <u>***User Understanding and Personalization***</u>**:** Different users have different kind of interests and it is highly likely that user is searching from his field of interest or work. Milad Shokouhi in [2] has discussed use of user's short and long history as well as other demographic features of location, user's age and gender, and region features for query auto completion, but also gives the idea of using the methodology for intent prediction too. This has been described in *Section 2.2.3*.

Figure 1.7 shows an example for the difference in results that can be brought by search with deep understanding from basic search. The query *"bing features"* is put in Quora and to restrict the domain of search to Quora, the same query has been refined as *"bing features quora"* being put in Bing, which suggests that the intent must have been to search for various features in Bing. Based on results obtained, assuming that Bing is doing deep understanding of the query, whereas Quora is not, the difference can easily be judged.

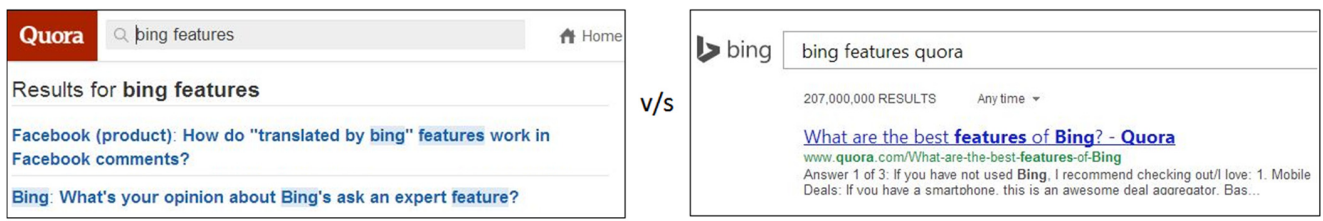

*Figure 1.7: An example for results variation by Advances in Search*

# 2 PHASES IN DEEP QUERY UNDERSTANDING

Query Understanding module has to primarily deal with three purposes – infer the intent of the query, guide him towards a precise intent by providing suggestions and refine the query for retrieval of better results. Figure 2.1 shows the components and their position to handle above mentioned purposes. The orange portion composes modules for the task of Query Inference, green portion for Query Refinement and blue portion for Query Suggestion.

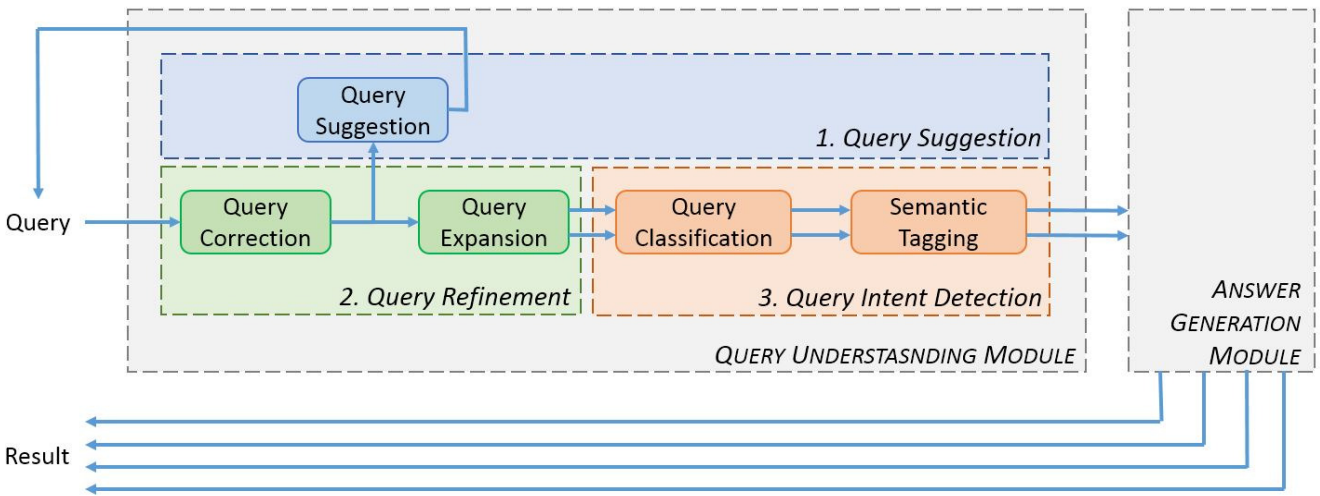

*Figure 2.1: Query Understanding Module Broad Components*

When a query arrives to query understanding module, it is first dealt by *Query Refinement* sub-module. The Query Correction component of which, processes the query with spelling correction, acronym expansion, word splitting, words merger and phrase segmentation. The corrected query is passed to *Query Suggestion* sub-module that discovers and suggest more queries that could lead to retrieval of better results. After one or more cycle of iterative procedure of *"Query Correction → Query Suggestion → User Query"*, the enhanced query is passed on to Query Expansion component, that adds some more synonymy similar queries to reduce the chances of result document miss due to term mismatch. This set of queries is then passed to *Query Intent Detection* sub-module which infers the precise intent of the query. The Query Classification component identifies one or more *segments* that are targeted by the query. A *Segment* is defined as one of the different categories like sports, weather, travel or video etc. e.g. query *"brazil germany"* can fall in sports segment (because of some football match) as well as travel segment (because of the search of flights from Brazil to Germany). Then the query is passed to the Semantic Tagging component for precise *intent detection.* The expanded set of intent tagged queries is then passed by the Query Understanding Module to the Answer Generation Module which then retrieves the documents based on its index.

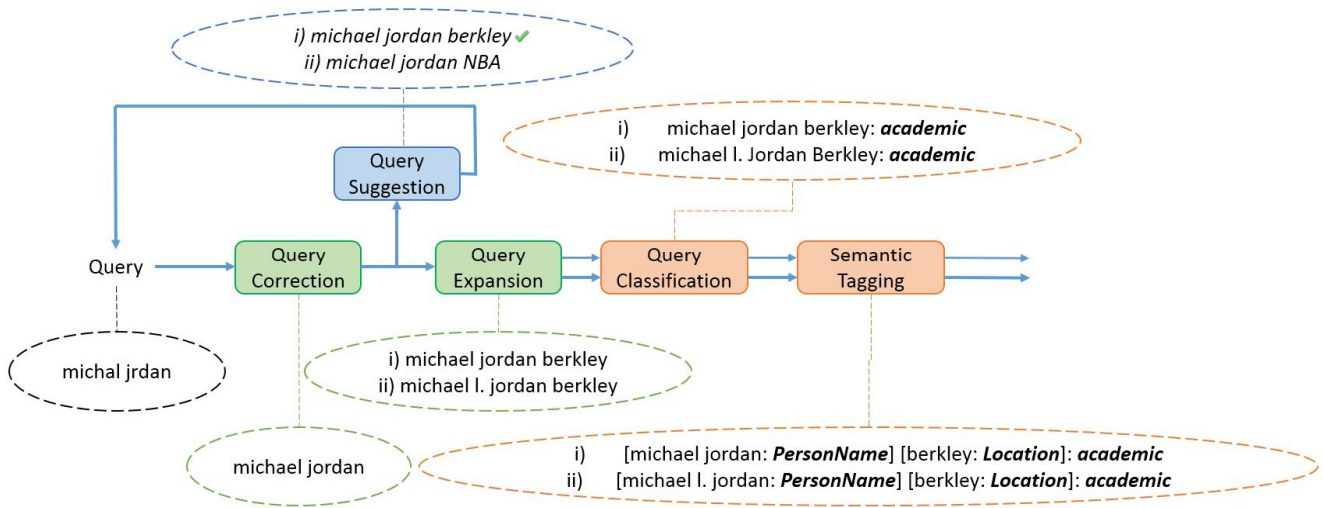

*Figure 2.2: Example of task done at each component*

Figure 2.2. takes an example seed query "michal jordan", and shows the various operations done at each component. Note that the seed query is misspelled. Query Correction component corrects it, Query Suggestion component suggests the likely queries that can be put, from which suppose that the user selects the query *"Michael Jordan berkley",* then the Query Expansion modules finds more similar synonymous queries to reduce the chances of result miss due to term mismatch. Then the Query Classification component classifies the obtained queries in one of the target categories. Finally Sematic Tagging is done to precisely identify the intent. The recent methodologies of each component has been discussed in the following sections.

## 2.1  QUERY CORRECTION (USING A MODIFIED CRF: CRF-QR)

Query correction module is primarily responsible for reformulating the ill-formed and mistaken search queries which will lead to retrieval of better results. The process includes operations like spelling error correction, splitting of a word into two for mistakenly merged terms, merging of two terms that actually signifies a single word but mistakenly are space separated, phrase segmentation, word stemming for better matching in documents, and acronym expansion which means expansion of terms like 'CSE' to 'Computer Science and Engineering' terms.

To accomplish the task of query refinement, Jiafeng et. al. in [10] has proposed an advanced model of Conditional Random Field (CRF), which they refer as CRF-QR. "*CRF model is discriminative graphical model, which focuses on modeling the conditional distribution of unobserved state sequences given an observation sequence*" as stated by [6]. Since refinement of different terms of the query are often mutually dependent, they need to be addressed simultaneously. The use of CRF boosts the accuracy in such cases. E.g. query *"lectures on machne learn"* needs to be refined as *"lectures on machine learning"* after spelling correction of 'machne' and expansion of term 'learn' to 'learning' and then grouping the phrase "*machine learning*" together which leads to the better understanding of the intent.

Jiafeng et. al. in [10] have given an Extended CRF-QR model which are capable to handle the scenario when more than one refinements can be applied e.g. *'learm'* → *'learn'* → *'learning'*. For explanation purpose, the paper [10] has described a Basic CRF-QR that applies only one refinement task to a word.

### 2.1.1 Basic Model

As stated above, in this model it is assumed that given a query term, only one of the refinement operation can be applied in this model. Let $x = x_1 \, x_2 \ldots x_n$ is sequence of original query words, and let $y = y_1 \, y_2 \ldots y_n$ is the corresponding sequence of refined query words, where $n$ denotes the length of the sequence, assumed to be of same length.

If the $y_i$ had depended only on xi we could have used a simple conditional probability model $\Pr(y_i|x_i)$ for each of the $x_i$. But, as described above the $y_i$'s could be mutually dependent on each other and all the $x_i$'s i.e. complete sequence $x$, and the hence $y_i$ should be conditioned on other $y_i$'s and all the $x_i$'s. Hence now the query refinement task can be re-defined as finding the sequence $y^*$ such that $y^* = \arg max_y \Pr(y|x)$ is satisfied, where CRF model should have been used to calculate the conditional probability $\Pr(y|x)$, which is quite intractable and will require very large amount of data because $\Pr(y|x)$ means the probability of occurrence of each valid word will have to be checked for a given $x_i$.

To reduce the space of y for a given x, the paper has extended the CRF model and formed CRF-QR model. The key idea of CRF-QR model is to condition y (the refined term) on operation of refinement also, along with the $x$ (original term) i.e. instead of calculating $Pr(y|x)$, the CRF-QR model calculates the $Pr(y, o \mid x)$, where $o = o_1 \, o_2 \, \ldots \, o_n$ is the sequence of operations to transform $x$ to $y$. Each operation $o_i$ is the one performed to transform $x_i$ into $y_i$. The operations are like deletion, insertion or substitution for Spelling Correction. Similar operations are defined for Word Splitting, Word Merging, Phrase Segmentation, Word Stemming and Acronym Expansion. So, now the mapping from *x's* to *y's* will not be completely free, and are restricted drastically be operations. Figure 2.3, taken from [10], depicts the CRF-QR in the graphical model, in which vertexes (*y's*) denote refined terms, edges denote dependencies and there are conditional vertexes of query words and operations. Note that each $y_i$ is conditioned on *x* (*for representation edge has been made to corresponding $x_i$ only*), refinement operation and adjacent refined terms. The adjacent operations ($o_i$'s) are not mutually conditioned in the model as that is implicitly incorporated due to mutual conditioning of $y_i$'s.

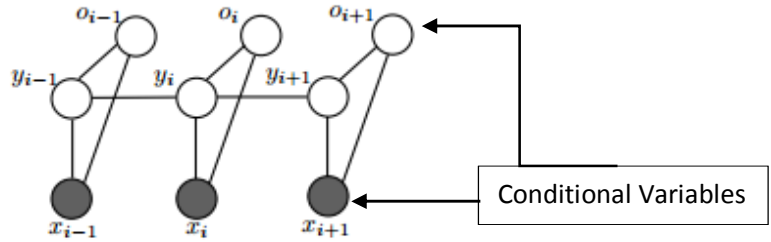

Figure 2.3: Basic CRF-QR Model

Hence, the query refinement problem is formulated as the task of getting the sequence of y*, such that it satisfies $y^*o^* = \arg max_{y,o} \Pr(y, o|x)$.

The paper states that, "*in general a graphical model can be written as a product of potential functions over the maximal cliques of the graph, where a potential function is strictly positive and real valued*". The CRF-QR model contains two types of maximal cliques in its graphical representation. One is the edge between $y_{i-1}$ and $y_i$, $i = 1, 2, ..n$ for which let $\phi(y_{i-1}, y_i)$ be the potential function and the second one is the edge between $y_i$ and $o_i$ conditioned on $x$, $i = 1, 2, ... n$ for which let $\phi(y_i, o_i, x_i)$ be the potential function. So, the model can be described by the eqn. (2.1) where $Z$ is the normalizing factor.

$$\Pr(y, o|x) = \frac{1}{Z(x)} \prod_{i=1}^{n} \phi(y_{i-1}, y_i) \phi(y_i, o_i, x_i) \tag{2.1}$$

The two potential functions are assumed to be of exponential nature and hence can be represented as eqn. (2.2) and eqn. (2.3) respectively.

$$\phi(y_{i-1}, y_i) = \exp\left(\sum_k \lambda_k f_k(y_{i-1}, y_i)\right) \tag{2.2}$$

$$\phi(y_i, o_i, x_i) = \exp\left(\sum_k \lambda_k h_k(y_i, o_i, x)\right) \qquad (2.3)$$

These equations capture the notion that the refined term $y_i$ depends on the operation ($o_i$) at that position and the entire sequence $x$. Therefore, in CRF-QR $h(y_i, o_i, x)$ and $f(y_{i-1}, y_i)$ are used as features to represent the dependency relationship.

### 2.1.2 Learning and Prediction

A labeled dataset of $(x^{(1)}, y^{(1)}, o^{(1)}), \ldots, (x^{(N)}, y^{(N)}, o^{(N)})$ is used to train the CRF-QR, where $x^{(i)}$ is the user query and $y^{(i)}$ is the corresponding refined query. The training means estimating the parameter $\hat{\lambda}$ by maximizing the regularized log-likelihood function given in eqn. (2.4) for the training data with respect to the model.

$$\hat{\lambda} = \arg max_\lambda \left\{\sum_{i=1}^{N} \log(Pr_\lambda(y^{(i)}, o^{(i)}|x^{(i)})) - C||\lambda||_2\right\} \qquad (2.4)$$

In eqn. (2.4), C is the coefficient and $||.||_2$ means $L_2$ norm. Note that the log-likelihood function is convex and hence a global maxima is bound to exist.

$$y^* o^* = \arg max_{y,o} \Pr(y, o|x) \qquad (2.5)$$

For prediction purpose, for the given query x, *Viterbi algorithm* is used to get the most likely refined query $y^*$ satisfying eqn. (2.5). "*Viterbi algorithm is a dynamic programming algorithm for finding the most likely sequence of hidden states – called the Viterbi path – that results in a sequence of observed events.*"

### 2.1.3 Features

The functions $h(y_i, o_i, x)$ and $f(y_{i-1}, y_i)$ represents the feature vector and need to be defined. Note that $f(y_{i-1}, y_i)$ represents the relation among adjacent words $y_{i-1}$ and $y_i$ of the refined query. This featured can be defined by the function as in eqn. (2.6)

$$f(y_{i-1}, y_i) = \log \Pr(y_{i-1}|y_i) \qquad (2.6)$$

$\Pr(y_{i-1}|y_i)$ is the conditional probability of occurrence of $y_i$ after $y_{i-1}$ in a corpus. This probability can be obtained by counting the frequency of bigrams in any sufficiently big corpus or query log data.

The second feature $h(y_i, o_i, x)$ represents the refinement operation $o_i$ that is used to transform $x_i$ into $y_i$, conditioned on $x$. Hence, this function can be straightforward defined by the eqn. (2.7).

$$h(y_i, o_i, x) = \begin{cases} 1, & \text{if } y_i \text{ is obtained from } x_i \text{ after operation } o_i \\ 0, & \text{otherwise} \end{cases} \quad (2.7)$$

### 2.1.4 Extended Model

The purpose of extended model is to refine those queries too that have more than one mistakes i.e. from given term $x_i$, there needs to be more than one operations $\vec{o} = o_{i,1}, o_{i,2}, \ldots o_{i,n}$ i.e. at position $i$, the $n$ operations have been done to obtain $y_i$. The intermediate results are shown by $\vec{z_i} = z_{i,1} z_{i,2} \ldots z_{i,m-1}$. Let $z_{i,0}$ denotes the starting query term $x_i$ and $z_{i,m}$ denotes the refined query term $y_i$.

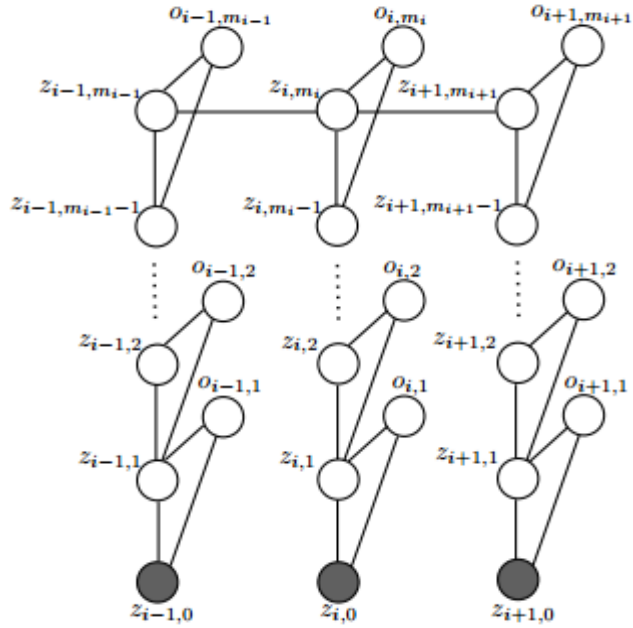

*Figure 2.4: Graphical representation of Extended CRF-QR*

Figure 2.4 from [10], shows the graphical representation of the extended CRF-QR model. Note that only the final (refined) terms have been mutually conditioned, and each refinement is conditioned on the term obtained by previous operation. As earlier in basic model, again the conditional probability distribution can be obtained by defining the potential functions on maximal clique and is given by the eqn. (2.8).

$$\Pr(y, \vec{o}, \vec{z} | x) = \frac{1}{Z(x)} \prod_{i=1}^{n} (\phi(y_{i-1}, y_i) \prod_{j_i=1}^{m_i} \phi(z_{i,j_i}, o_{i,j_i}, z_{i,j_i-1}) \quad (2.8)$$

Here again the potential functions have been assumed to be of exponential nature and so can be represented as eqn. (2.9) (which is same as eqn. (2.2)) and eqn. (2.10) respectively.

$$\phi(y_{i-1}, y_i) = \exp(\sum_k \lambda_k f_k(y_{i-1}, y_i)) \tag{2.9}$$

$$\phi(z_{i,j_i}, o_{i,j_i}, z_{i,j_{i-1}}) = \exp(\sum_k \lambda_k h_k(z_{i,j_i}, o_{i,j_i}, z_{i,j_{i-1}})) \tag{2.10}$$

Similar to Basic Model, here also the features can be defined, and can be used for learning and prediction. So, by using the extended CRF-QR model, the query correction module will be able to refine the queries requiring multiple intermediate operations too.

The comparative experiments done by the paper shows that the relative improvement brought by CRF-QR is 2.26% and 1.21% in terms of F1 score and accuracy as compared to best of baseline method.

## 2.2 QUERY SUGGESTION

Query suggestion means suggesting similar queries to the user, to guide him to formulate a query with precise intent. The section discusses the work done by Huanhuan et. al. in [7] which has given an approach by mining Click-Through and Session Data, the work done by Allan et. al in [9] which identifies the interleaved tasks and recommends queries accordingly, and Milad et. al in [2] which has personalized the suggestions based on lot many user's personal features. The papers have discussed and used the approaches that can be used for intent detection also, such as using context for resolving ambiguity in intent e.g. given a query *"gladiator"*, it is difficult to infer whether the intent is to get history of gladiator or the movie gladiator. But if the previous query was *"a beautiful mind"* which is a query from domain of movie concept, it is highly probable that the current query of *"gladiator"* is also for the movie intent.

### 2.2.1 Context aware Query Suggestion

Huanhuan et. al. in [7] has defined a data structure *Concept Suffix Tree,* which has vertices representing the state after transition through different concepts from beginning of session. E.g. Figure 2.5, taken from [7], represents a concept suffix tree, which shows that from beginning of session, the first concepts moved upon are *C1, C2, C3, C4* and *C5* and the label $C4C3C1$ represents the state after the transition *Beginning* → *C1* → *C3* → *C4*. The key idea in this paper is to build a *Concept Suffix Tree* from available query log. For mapping query to concept, the paper has described a clustering method to discover different concepts. Once the tree has been built, query suggestions are made by transiting on this tree, according to observed concept of each query. At any given point, the high frequency queries of the arrived state are suggested.

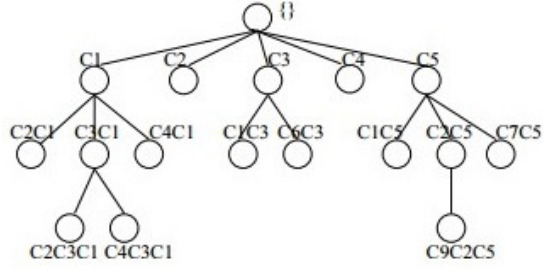

*Figure 2.5: A Concept Suffix Tree*

**Concept Discovery:** Queries are clustered according to the similarity of its set of clicked URLs and each of the identified cluster is considered as a concept. The set of URLs clicked for the queries is known as its feature set being created using the raw search log. To create the feature set of each query, a bipartite graph is created in which one part contains nodes for each unique query and another part has nodes for each unique URL and an edge $e_{ij}$ is created between query node $q_i$ and URL node $u_j$, if $u_j$ is a clicked URL for $q_i$, with edge weight being the count for number of times $u_j$ is a click of $q_i$. Such a bipartite is known as click-through bipartite. Figure 2.6, taken from [7], shows a sample where the bipartite graph has 4 query nodes and 5 URL nodes, and for an instance for query $q_1$, URL $u_1$ has been clicked 30 times and the URL $u_3$ has been clicked 20 times.

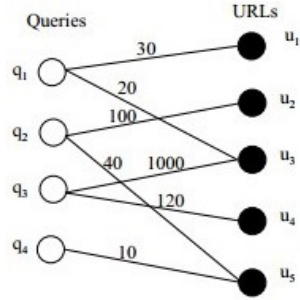

*Figure 2.6: An example of Click-Through bipartite*

Now, each query $q_i$ is represented as an $L_2$-normalized vector, where each URL in the bipartite corresponds to one dimension. For the feature vector of query $q_i$, the $j^{th}$ element is given by eqn. (2.11)

$$\vec{q_i} = \begin{cases} norm(w_{ij}) & if\ edge\ e_{ij}\ exists \\ 0 & otherwise \end{cases} \quad (2.11)$$

where norm is defined by eqn. (2.12)

$$norm(w_{ij}) = \frac{w_{ij}}{\sqrt{\sum_{\forall e_{ik}} w_{ik}^2}} \quad (2.12)$$

Based on the defined feature vector, the distance between two queries $q_i$ and $q_j$ can be defined by the following eqn. (2.13)

$$distance(q_i, q_j) = \sqrt{\sum_{u_k \in U}(\vec{q_i}[k] - \vec{q_j}[k])^2} \qquad (2.13)$$

Now using this defined distance equation between any two queries, paper has described an incremental clustering method to find the clusters, where each cluster will represent a discovered concept. Both the mappings *query → cluster* and *cluster → query* are maintained (in form of dictionary) for their application in the online query suggestion phase, as described below in respective section. The clustering method has been described below.

***Clustering Method:*** Since each query feature vector contains an element for each of the URL, the dataset will become extremely high dimensional. Therefore, there is an absolute need to tackle the *"curse of dimensionality"*. Also the search log will increase dynamically, so clustering needs to be maintained incrementally.

To describe the clustering method, following definitions are required to be known. The normalized centroid for any cluster is given by the eqn. (2.14)

$$\vec{c} = norm(\frac{\sum_{q_i \in C} \vec{q_i}}{|C|}) \qquad (2.14)$$

where |C| is the number of queries in the considered cluster C. The distance between any query of the cluster and the centroid is given by eqn. (2.15)

$$distance(q, C) = \sqrt{\sum_{u_k \in U}(\vec{q}[k] - \vec{c}[k])^2} \qquad (2.15)$$

and to evaluate the compactness of clusters, diameter parameter is defined as eqn. (2.16)

$$D = \sqrt{\frac{\sum_{i=1}^{|C|} \sum_{j=1}^{|C|}(\vec{q_i} - \vec{q_j})^2}{|C|(|C| - 1)}} \qquad (2.16)$$

An upper threshold on $D_{max}$ is set on $D$ to control the granularity of clusters.

Initially there is no cluster and the first query is treated as the first cluster. Further for each query $q_i$, the closest cluster ***C*** is found, and then the diameter of $C \cup \{q\}$ is tested. If the diameter is less than $D_{max}$, the cluster $C$ is updated to $C \cup \{q\}$, otherwise a new cluster containing only $\{q\}$ is created. This method requires only one scan through the query log.

A potential problem here is that finding the closest cluster might be time consuming. To tackle this problem, the paper has defined a set $Q_q$ as set of queries which share at least one URL with $q$, and state the key idea that to find the closest cluster to $q$, only those clusters needs to be checked which contain at least one query in $Q_q$. According to some experimental analysis, a query is connected to an average number of 8.2 URLs, and average degree of URL nodes is only 1.8. So, the size of $Q_q$ comes equal to 8.2 * (1.8 - 1) = 6.56. So, on an average only 6.56 clusters will have to be checked for each query.

***Building Concept Suffix Tree*** *(method is similar to Trie data structure):* Each session in the query log has a sequence of queries $q_s = q_1 q_2 \ldots q_l$. From above concept discovery, one to one mapping from each query to corresponding concept has been defined. With each query $q_i$ in the sequence, corresponding sequence of concepts $c_s = c_1 c_2 \ldots c_l$ can be built. But often consecutive queries are similar and liable to fall in same cluster, hence for such a scenario the concept is recorded only once. The Concept Suffix tree is built while finding the next concept in the concept sequence ($c_s$) for each session. Note that, root node represents empty sequence, and each session starts at this node. Now, at any state, if the next concept found is child of current concept node then move to that child node, otherwise create the child node with that concept and move to it.

***Online Query Suggestion:*** The built concept suffix tree is used to suggest new queries. The queries made in a session are tracked and the mapping $q_i \rightarrow c_i$ (dictionary) which was built earlier, is used to traverse the suffix tree with each query. This might happen that the query entered is new and there does not exist any mapping corresponding to it. In this case, the state is restored to the root node. To suggest new queries, the high frequency queries of current node are returned using the mapping $c_i \rightarrow q_i$ (dictionary).

### 2.2.2 Task aware Query Suggestion

As defined earlier in Section 1.2.3, the paper [9] has given the definition of *On-Task Queries* as those queries which are of the same task as that of the *Reference Query* (current query entered) and *Off-Task Queries* which are of different task as that of the *Reference Query*. The paper showed through experiments that while considering contexts as done in Section 2.2.1, if Off-Task Queries are considered then it adversely affect quality measured using Mean Reciprocal Rank (*MRR*). Figure 2.7 from [9], shows results of considering on-task and off-task queries versus considering only the reference query, which conclude that considering off task queries in the context does not help and even harm the quality of recommendation as compared to the results of considering only the reference query.

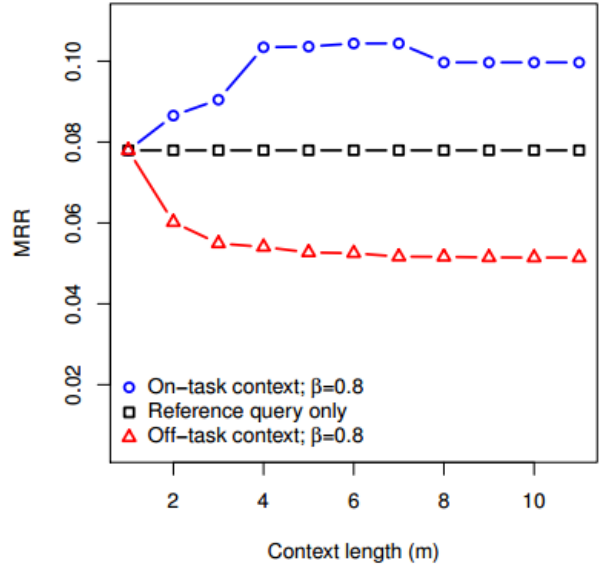

*Figure 2.7: The effect of adding on-task and off-task queries versus only the reference query on recommendation MRR.*

Hence it becomes important to identify the *On-Task* and *Off-Task* queries and accordingly consider as context for recommendation.

The paper has discussed an approach for automatic search task identification. To identify whether two queries $q_i$ and $q_j$ are part of the same task or not, the paper has defined the following two similarity measures:

i) *Lexical score ($s_{lexical}$):* average of the *Jaccard coefficient between trigram extracted from the two queries* and *(1 – Levenshtein edit distance)*. This score will fall in [0,1]. The lexically very similar queries like the ones where only one term is added / removed / reordered, or spell corrected will have this score close to 1.

ii) *Semantic score ($s_{semantic}$):* This function requires two values: one is $s_{wikipediea}(q_i, q_j)$, that first creates the vectors $v_i$ and $v_j$ which consists the tf·idf scores of every Wikipedia document with respect to $q_i$ and $q_j$, respectively and then returns the cosine similarity between $v_i$ and $v_j$. The second function is $s_{wikitionary}(q_i, q_j)$ which is computed similarly, but over Wiktionary entries. Using these two, the $s_{semantic}$ is computed as given in eqn. (2.17)

$$s_{semantic}(q_i, q_j) = \max(s_{wikipediea}(q_i, q_j), s_{wikitionary}(q_i, q_j)) \qquad (2.17)$$

Using the two above defined scores, the combined score is defined as in eqn. (2.18)

$$s(q_i, q_j) = \alpha \cdot s_{lexical}(q_i, q_j) + (1 - \alpha) \cdot s_{semantic}(q_i, q_j)) \qquad (2.18)$$

With this defined similarity function, same-task scoring of a context (session) *C,* as given in the following eqn. (2.19).

$$sametask(i, j, C) = s(C[i], C[j]) \qquad (2.19)$$

where *C[i]* and *C[j]* are two queries of the considered context. This value above a threshold for this score will identify On-Task queries.

### 2.2.3 Personalization in Query Suggestion

The above two sections discussed the approach for query suggestion considering the context but there was no personalization based on user specific parameters like location, age, gender etc. The paper [2] has given an experimentally found example that for the hit character 'i', the query term 'instagram' is more popular among young female below the age of 25, and in contrast 'imdb' is more popular among male users, particularly between the age of 25 to 44. Based on such studies, the paper [2] has proposed and discussed the effect of several user-specific and demographic-based features. The paper has targeted the use of these features for query auto-completion but these can be used for query suggestion also, as these two tasks are closely related. There are primarily 4 types of Feature Groups – *i) Short History Features, ii) Long History Features, iii) Demographics, and iv) Most Popular Completion* as given in Table 2.1, taken from [2].

*Table 2.1: Description of Features used for Personalization*

| | Feature | Description | Change in MRR |
|---|---|---|---|
| Short history | PrevQueryNgramSim | n-gram similarity with the previous query in the session (n = 3). | +0.91% |
| Short history | AvgSessionNgramSim | Average n-gram similarity with all previous queries in the session (n = 3). | |
| Long history | LongHistoryFreq | The number of times a candidate is issued as query by the user in the past. | +5.57% |
| | LongHistorySim | Average n-gram similarity with all previous queries in the user's search history. | |
| Demographics | SameAgeFrequency | Candidate frequency over queries submitted by users in the same age group. | +3.8% |
| | SameAgeLikelihood | Candidate likelihood over queries submitted by users in the same age group. | |
| | SameGenderFrequency | Candidate frequency over queries submitted by users in the same gender group. | +3.59% |
| | SameGenderLikelihood | Candidate likelihood over queries submitted by users in the same gender group. | |
| | SameRegionFrequency | Candidate frequency over queries submitted by users in the same region group. | +4.58% |
| | SameRegionLikelihood | Candidate likelihood over queries submitted by users in the same region group. | |
| MPC | SameOriginalPosition | The position of candidate in the MPC ranked list. | These features were included with all other features |
| | SameOriginalScore | The score of candidate in the MPC ranked list computed based on past popularity. | |

Overall, by consideration of all the user specific and demographic features, there was an increment of 9.42% in terms of MRR for query recommendation. The most influential feature came out to be demographic features such as location (region), which can be seen as the queries based on the interest of local people, which can be affected by some local event or local conditions.

## 2.3 QUERY EXPANSION

Query expansion is the process of reformulating the actual query (seed) to tackle the challenge of term-miss in match. The terms of the seed query are expanded with other similar terms to match additional documents. The query expansion technique involves:

*i)* Finding synonyms of words, and searching for the synonyms as well.

*ii)* Finding all the various morphological forms of words by stemming.

*iii)* Re-weighting the terms in the original query, which means to describe the importance of the terms. The document having higher occurrence of more important terms will have higher ranking.

The naïve approach for above tasks can be performed by exhaustive lookup in a thesaurus. But in recent advancements, more efficient techniques have been developed to make the system learn itself which words can be expanded and to which. The paper [11] has given an approach to exploit search logs for Query Expansion, based on path-constrained random walks. The methodology builds a labeled directed graph using the search log data of <query, document> pairs, known as click-through data. There are three type of nodes – Queries, Documents and Words.

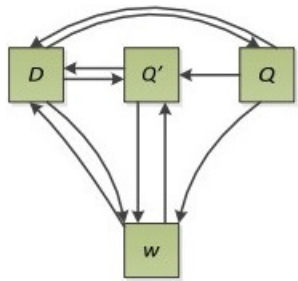

*Figure 2.8: Example of graph built using Search Log*

Figure 2.8 from [11] shows an example, in which the node *Q* represents the seed query, the nodes represented by *Q'* are queries present in the search log, nodes *D* are the documents and nodes *W* that occur in queries and documents. The word nodes are candidate expansion terms. The edges are labelled with the *relation r*, each of which is having a scoring functions, described by Jianfeng Gao et. al. in [11] based on factors like word occurrences in document, set of documents clicked for queries etc. The value obtained by scoring function represents the probability of transition from source node to target node in one step random walk on that edge, denoted as $P(t|s,\theta_r)$.

Using the definitions of scoring functions for each edge $score_{\theta_r}(s \rightarrow t)$, where $s \rightarrow t$ represents a directed edge from $s$ to $t$, the probability of transition over a given edge is given by eqn. (2.20)

$$P(t|s,\theta_r) = \frac{\exp(score_{\theta_r}(s \rightarrow t))}{\sum_{t_i} \exp(score_{\theta_r}(s \rightarrow t_i))} \tag{2.20}$$

Now, for the constructed graph, any path $\pi$, obtained through random walk, that starts from input query node $Q$ and ends at a word node $w$ gives the probability of picking $w$ as an expansion term i.e. $P(w|Q,\pi)$. This probability is given by product of probabilities on all the edges encountered on path $\pi$.

## 2.4 QUERY CLASSIFICATION

Query classification is the task of classifying a given query in one of the predefined class, generally given as a taxonomy. A taxonomy is a tree of categories in which each node corresponds to a predefined category. Figure 2.8 from [6], shows an example of Taxonomy. Each of the query can be put in one or more of the leaf node(s).

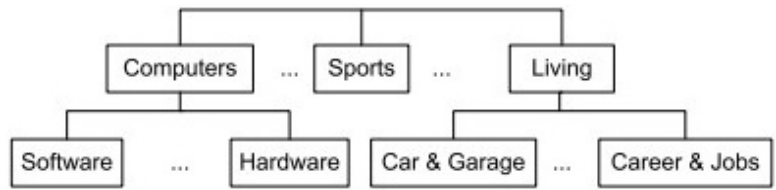

*Figure 2.9: An example of Taxonomy*

The sequence of labels of the nodes from root to the leaf node, defines the semantic meaning of the category at that leaf node. Classification of query is more challenging than document text classification because of the two reasons – i) The short length of the web queries and ii) many queries are ambiguous. The task of query classification is important for the precise intent detection.

### 2.4.1 Context aware Query Classification

An example for context aware classification could be for the query *"Michael Jordan"*. This query can fall in the intent of *Sports* category representing the search of famous Basketball player or the intent of *Academics* representing the search for famous Machine Learning professor. If the context can be identified to be of Sports then the current query is more probable to be of *Sports* intent. The paper [6] has discussed the use of context for classifying the query with modelling on *Conditional Random Field (CRF)*.

*"CRF model is discriminative graphical model, which focuses on modeling the conditional distribution of unobserved state sequences given an observation sequence"* as stated by [6]. The CRF has the strength to process sequential data and that it can incorporate rich features.

*Features of the CRF Model:* The paper has given two kinds of features

i) *Local Features*: that do not consider any context information. Three such features have been defined as following:

   a. *Query Terms:* Each individual term of the query is taken as feature to support the category label. But this requires a huge amount of training data to capture all pairs of valid query term $q_t$ and the target category label $c_t$. So, this feature will not work for a new query where some or all the query terms do not occur in the training data.

   b. *Pseudo Feedback:* This feature reflects the confidence that $q_t$ is actually in category $c_t$. To compute this feature, a query $q_t$ with concept $c_t$ is submitted to an external web directory. The category of each of the top M results are obtained in domain of categories of target taxonomy. Then, a general label confidence score is defined as in eqn. (2.21)

   $$GConf(c_t, q_t) = \frac{M_{c_t, q_t}}{M} \qquad (2.21)$$

   where $M_{c_t, q_t}$ is the count of returned search results for which category label is $c_t$. This score represents the feature.

   c. *Implicit Feedback:* This feature is computed similar to the Pseudo Feedback but for this feature, only the clicked URLs by the user are considered instead of the top-M results.

ii) *Contextual features*: These features consider context information and are required to reflect the association between the adjacent category labels. Two such features have been defined as following:

   a. *Direct association between adjacent labels:* The association between the adjacent labels can be captured by the number of occurrences of the pair of adjacent labels $<c_{t-1}, c_t>$, where $c_{t-1}$ and $c_t$ are leaf categories in the target taxonomy. The higher weight of $<c_{t-1}, c_t>$ shows the higher probability of transit from $c_{t-1}$ to $c_t$. The weights for this feature are learned during the training process of CRF.

   b. *Taxonomy-based association between adjacent labels:* Due to the limitation of training data, many transition between the two categories may not occur in the training data. Also, this is not necessary that number of observed transitions reflect the real world distribution. In such a case, the CRF will not be able to properly capture the direct association between categories. To deal with this problem, structure of taxonomy is also considered along with feature of direct association between adjacent labels. It is quite intuitive that association between two sibling categories should be stronger than that of two non-sibling categories. So, for a given pair of adjacent labels $< c_{t-1}, c_t >$ where both the categories are at level n, n-1 features of taxonomy-based association between $c_{t-1}, c_t$ are considered as $\{< \alpha^i_{c_{t-1}}, \alpha^i_{c_t} >\}$ for each i ∈ $[1, n-1]$. These weights of these features are learned during the training of CRF model.

The discussed features came out to be quite promising as the paper has done comparative experiments which conclude that the overall F1-score increased to 8.20 x $10^{-3}$ from 8.64 x $10^{-4}$ that was with basic features without considering context.

## 2.5 SEMANTIC TAGGING

Semantic Tagging identifies the semantic concepts of a word or phrase. Semantic Tagging in query phrases enhances the retrieval when the documents are also semantically tagged, in which case the semantic tags will also be matched along with the query terms. Figure 2.2 shows an example that clarifies what is done under sematic tagging. In that example the phrases *"michael jordan"* was tagged to be *"PersonName"* and *"berkley"* was tagged to be "Location". These attributes *"PersonName"* and *"Location"* are known as semantic tags.

For semantic tagging the commonly used features are n-grams, regular expressions, lexicons and transit features. Some more advancement introduced the use of shallow parsing features like *Part of Speech (POS)* tags, which worked very well for short-phrased segments like *Named Entities*. But as stated in *Section 1.1*, use of natural language queries is increasing, and in general these natural language queries are longer than the keyword represented queries. For such longer and grammatically correct queries, *Sentence Level Long Segments* are also present and that can't be identified by shallow parsing. An example for sentence level long segments, given by [5] is as following in Figure 2.10, where a long phrase has been tagged with the semantic tag *"<Plot>"*.

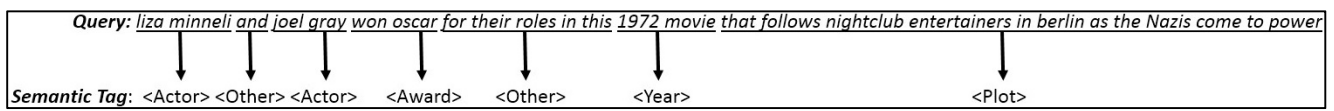

*Figure 2.10: Example of Sentence Level Long Segments in Semantic Tagging - "<Plot>"*

To capture such sophisticated semantic tags, the recent techniques leverages more advanced features like *Hierarchical Syntactic Features* and *Semantic Dependency Features* as discussed in the paper [5], described in the following sub-section.

### 2.5.1 Using NLP Technique - Hierarchical Parsing Structures

The author Jingjing Liu et.al in [5] have proposed to use semi-Markov CRFs. The problem is defined as given a sequence of words $x = x_1, x_2, \ldots, x_M$, another sequence $s = s_1, s_2, \ldots, s_N$ which represents the N segments made over M words $N \leq M$, and each $s_i$ represents the classification of the $i^{th}$ segment. The tuple $s_j = (u_j, v_j, y_j)$ can represent the corresponding segment where $u_j$ and $v_j$ are start and end indices of the segment and $y_j$ is the class label. With this notion, segmentation and classification of each segment can be modeled by eqn. (2.22)

$$p(s|x) = \frac{1}{Z_\lambda(x)} \exp\left\{\sum_{j=1}^{N+1} \lambda \cdot f(s_{j-1}, s_j, x)\right\} \quad (2.22)$$

where $f(s_{j-1}, s_j, x)$ is a vector of feature functions defined on the segment level. For training purpose, $\lambda$ is estimated to maximize the likelihood of observed training data while regularizing model parameters. Using the trained model, for any given sequence $x$, the sequence $s$ can be predicted. Author has defined the following vector of feature functions for training purpose are defined as following:

i) *Hierarchical Syntactic Features:* These features are used to capture the syntactic characteristics in a query through parsing structures like sub-clauses in the parse tree generated e.g. for a movie segment query (*Note that Query Classification has already been done before Semantic Tagging*), occurrences of consecutive NN supports the notion of <Actor> tag. More interesting thing to be noted is that SBAR (*subordinate clause*) field is likely to be a description that can be inferred as <Plot> tag. With this notion, one of the feature is described below:

*Node Feature*: This feature captures the co-occurrence of a semantic tag and the root node of the segment being considered. This is represented by eqn. (2.23)

$$f(s_{j-1}, s_j, x) = \delta(N(s_j) = t)\delta(y_j = b) \quad (2.23)$$

where $N(s_j)$ represent root of the sub-tree of segment $s_j$, $t$ is any parse tree node and $y_j$ is a class label, $b$ could be any semantic tag.

Other features of this category are *Node-And-Length Feature, Node-and-Children Feature, Node-and-POS Feature, Ancestors-and-Length Features, Node-and-Word-Before*

*Feature* and *Node-and-Phrase-Before Feature*. Table 2.3 clarifies meaning of each by showing the value of these features for the segment "*prominent drug usages and aliens*", which is labeled as <Plot>. The actual query *Q* is "*a 1984 cult classic starring emilio estevez and harry dean stanton that features prominent drug usages and aliens*" and the corresponding linguistic parse tree is shown in Figure 2.11, taken from [5].

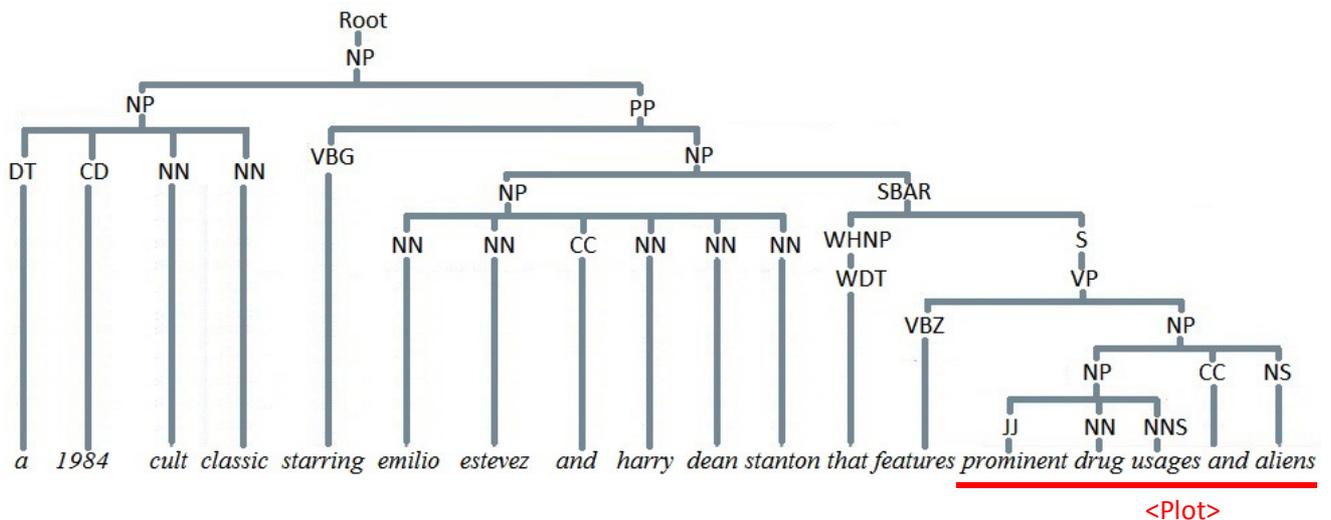

Figure 2.11: Linguistic Parse Tree for the Query *Q*

Table 2.2: Hierarchical Syntactic Features extracted for the Query *Q*.

| Feature | Example |
|---|---|
| Node | (NP) |
| Node-and-Length | (NP, 5) |
| Node-and-Children | (NP, NP-CC-NNS) |
| Node-and-POS | (NP, JJ-NN-NNS-CC-NNS) |
| Ancestors-and-Length | (ROOT-NP-SBAR-S-VP-NP, 6) |
| Node-and-Word-Before | (NP, "features") |
| Node-and-Phrase-Before | (NP, "that features") |

In Table 2.2, note that Node Feature is the root node (*NP*) in sub-tree (of linguistic sub-tree) corresponding to the considered segment. Node-and-Length feature is *(NP,5)* because there are 5 words in the segment, and so on for other features.

*ii)* *Semantic Dependency Features*: These features leverage the information about dependencies among different segments. E.g. consider the query *"show me a funny movie starring Johnny and featuring Carribbean Pirates"*, for which the verb *"featuring"* takes the arguments *"funny movie"* and *"Carribbean pirates"*. These dependencies are used as features as following:

a. *Solo-Dependency Feature*: This feature can be defined by eqn. (2.25)

$$f(s_{j-1}, s_j, x) = \delta(R(w) = r)\delta(arg_r(w) = s_j)\delta(y_j = b) \qquad (2.24)$$

where $R(w) = r$ means that word $w$ has a dependency relationship $r$, and $arg_r(w) = s_j$ means that one of the arguments of $w$ is the current segment $s_j$ i.e. there is one dependency from a word to a one of the semantic tag e.g. Verb(*featuring*), arg(<Plot>)

b. *Dual-Dependency Feature*: This feature is fired up when there is a dependency from a word to two of the semantic tags e.g. Verb(*featuring*), arg1(movie), arg2(<Plot>)

c. *Chain-Dependency*: This feature is fired up when there is a dependency like $w_1 \rightarrow w_2 \rightarrow$ *some semantic tag*. E.g. Show → movie → <Plot>

The paper has done experimental analysis which shows that the baseline method gave F1-score equal to 85.32%, whereas using the proposed dependency features the F1-score came to be 86.40%.

# 3 CONCLUSION AND FUTURE WORK

Query Understanding includes query intent detection, query suggestion and query refinement, out of which there is a lot of scope in the field of Query Intent Detection. With rise of personal digital assistants and expert systems, there is an increase in Natural Language Queries, and being grammatically correct, they can be leveraged to better understand the query and its intent. A lot many techniques like consideration of context, user's history and his location and other information like gender and age have already been incorporated to precisely detect the intent.

Another possible area of research could be to identify communities of users' based on the kind of queries being done by them. Then all the three tasks of Query Understanding Module can be tweaked for that community. E.g. suppose a pattern of queries is identified among few members of that community, which can be due to some event in their community, then that query can be suggested to other member of the community.

# REFERENCES

[1] Makoto P. Kato, Takehiro Yamamoto, Hiroaki Ohshima and Katsumi Tanaka, "Cognitive Search Intents Hidden Behind Queries: A User Study on Query Formulations," in *WWW Companion*, Seoul, Korea, 2014.

[2] Milad Shokouhi, "Learning to Personalize Query Auto-Completion," in *SIGIR*, Dublin, Ireland, 2013.

[3] Weotta, "Deep Search," 10 6 2014. [Online]. Available: http://streamhacker.com/2014/06/10/deep-search/. [Accessed 6 8 2014].

[4] W. Bruce Croft, Michael Bendersky, Hang Li and Gu Xu, "Query Understanding and Representation," *SIGIR Forum,* vol. 44, no. 2, pp. 48-53, 2010.

[5] Jingjing Liu, Panupong Pasupat, Yining Wang, Scott Cyphers and Jim Glass, "Query Understanding Enhanced by Hierarchical Parsing Structures," in *ASRU*, 2013.

[6] Huanhuan Cao, Derek Hao Hu, Dou Shen and Daxin Jiang, "Context-Aware Query Classification," in *SIGIR*, Boston, Massachusetts, USA, 2009.

[7] Huanhuan Cao, Daxin Jiang, Jian Pei, Qi He, Zhen Liao, Enhong Chen, Hang Li, "Context-Aware Query Suggestion by Mining Click-Through and Session Data," in *KDD*, Las Vegas, Nevada, USA, 2008.

[8] Zhen Liao, Yang Song, Li-wei He and Yalou Huang, "Evaluating the Effectiveness of Search Task Trails," in *WWW*, Lyon, France, 2012.

[9] Allan, Henry Feild and James, "Task-Aware Query Recommendation," in *SIGIR*, Dublin, Ireland, 2013.

[10] Jiafeng Guo, Gu Xu, Hang Li and Xueqi Cheng, "A Unified and Discriminative Model for Query Refinement," in *SIGIR*, Singapore, 2008.

[11] Jianfeng Gao, Gu Xu and Jinxi Xu, "Query Expansion Using Path-Constrained Random Walks," in *SIGIR*, Dublin, Ireland, 2013.